\documentclass[twocolumn,aps,superscriptaddress,showpacs,nofootinbib,floatfix]{revtex4}
\usepackage{epsfig,bm,feynmf}
\usepackage{graphicx}
\usepackage{amsmath}
\usepackage{dcolumn}

\usepackage{graphicx}
\usepackage[usenames,dvipsnames,svgnames,table]{xcolor}

\usepackage{tikz-feynman,contour}
\tikzfeynmanset{compat=1.0.0}

\begin{document}


\title{Bremsstrahlung photon from a hadronizing quark-gluon plasma}


\author{Taesoo Song}\email{t.song@gsi.de}
\affiliation{GSI Helmholtzzentrum f\"{u}r Schwerionenforschung GmbH, Planckstrasse 1, 64291 Darmstadt, Germany}




\begin{abstract}
Assuming that quark and antiquark numbers are separately conserved during hadronization, we calculate Bremsstrahlung photon from a hadronizing quark-gluon plasma.
The quark and antiquark numbers are obtained from the hadron numbers in the statistical model and the transition amplitudes for the hadronization from the constraint that all quarks and antiquarks must be consumed in the hadronization. 
Then Bremsstrahlung photon from the hadronization is obtained in the soft photon approximation, and we find that its contribution to the direct photon increases in low-energy heavy-ion collisions and in peripheral collisions where the lifetime of a quark-gluon plasma (QGP) is relatively short.
\end{abstract}


\maketitle

\section{introduction}
Ultra-relativistic heavy-ion collisions produce an extremely hot and dense nuclear matter which is possibly related to the state of the early universe.
Electromagnetic particles such as dilepton (virtual photon) and real photon are promising probe particles searching for the properties of the extreme matter. 
Since they have no strong charge, they do not interact with the produced matter but get through it without being interrupted.
Therefore, they deliver the information of the matter at their production sites and times~\cite{Linnyk:2015rco,David:2019wpt}.
  
The produced photons in heavy-ion collisions are classified into the decay photon and the direct photon.
The former is produed through the electromagnetic decay of hadrons, while the latter from the interactions of particles both in partonic and hadronic phases.
The direct photon is more interesting, because it discloses the properties of the matter.

Several years ago it has been measured at the RHIC and LHC that the elliptic flows of direct photon in heavy-ion collisions are comparable to those of pion and decay photon, which is called 'direct photon puzzle,' because the direct photon is continually produced from the initial stage where elliptic flows are not developed yet~\cite{Shen:2013vja,Paquet:2015lta,David:2019wpt,PHENIX:2014nkk,PHENIX:2015igl,ALICE:2015xmh,ALICE:2018dti}.
One possible way to explain the large elliptic flows is that direct photons are mainly produced in the late stage such as hadronic phase~\cite{Eggers:1995jq} rather than in partonic phase, and the out-of-equilibrium photon production might help it~\cite{Schafer:2021slz}. 

Production channels of direct photon in heavy-ion collisions are catogorized according to production stage.
The first one is the prouction before thermalization of the matter, which includes the primordial production and the pre-equilibrium one.
The former is the production through the scattering of partons in the colliding nucleons, which is calculable in pQCD.
Since the same photon is produced in p+p collisions, it can be scaled by the number of binary collisions in heavy-ion collisions.
The latter is presently unclear and depends on model for the pre-equilibrium matter~\cite{Berges:2017eom,Monnai:2022hfs,Monnai:2019vup,Churchill:2020uvk}.  

The second and third ones are, respectively, partonic and hadronic productions after the thermalization. 
The dominant channels in QGP are $q(\bar{q})+g\rightarrow q+\gamma$ and $q+\bar{q}\rightarrow g+\gamma$ while in the hadron gas phase $\pi+\pi\rightarrow \rho+\gamma$ and $\pi+\rho\rightarrow \pi+\gamma$ with $\pi$ and $\rho$ being changable to $K$ and $K^*$, respectively.

Other source of direct photon in both partonic and hadronic matter is Bremsstrahlung photon which is induced by the interactions of charged particles.
In QGP, for example, $q(\bar{q})+q(\bar{q})$ and $q(\bar{q})+g$ scatterings can produce Bremsstrahlung photon, 
because (anti)quark has nonzero electric charge.
  
Hadronization is a kind of interaction which confines free (anti)quarks into a bound state of hadron.
However, photon production from hadronization has barely been studied~\cite{vanHees:2014ida,vanHees:2011vb,Rapp:2013ema,Garcia-Montero:2019kjk,Fujii:2022hxa}. 
According to the lattice calculations the phase transition is crossover at small baryon chemical potential~\cite{Borsanyi:2010cj}, and the hadronization will be a smooth continous transition from the thermal distribution of free (anti)quarks to the thermal distribution of free hadrons~\cite{Song:2021mvc}.
Though it is not an instant interaction, the momentum changes of (anti)quarks through the hadronization will bring about the emission of Bremsstrahlung photon.
Since low-energy Bremsstrahlung photon is not affected by microscopic process but by macroscopic process, the incoming and outgoing momenta of (anti)quarks are the only necessary input to study the production of low-energy Bremsstrahlung photon~\cite{Song:2018wvd,Song:2022wil}. 

Hadronization happens in an extremely nonperturbative region of QCD and many things of it are not well known.
In this study we rely on the statistical model to obtain quark and antiquark number densities at $T_c$, assuming that quark and antiquark numbers are seperately conserved during the hadronization and they play the role of constituent quarks and constituent antiquarks of hadrons.
The transition amplitudes for hadronization are determined from the contraint that all quarks and antiquarks are consumed during the hadronization.


We first calculate the photon production from meson and baryon formations in section~\ref{photon}, which is applied to a QGP in section~\ref{QGP} to obtain the spectrum of Bremsstrahlung photon per unit volume of hadronizing QGP.
After comparing the results with experimental data, we give a summary in Section~\ref{summary}.

\section{Bremsstrahlung photon from meson/baryon formation}\label{photon}

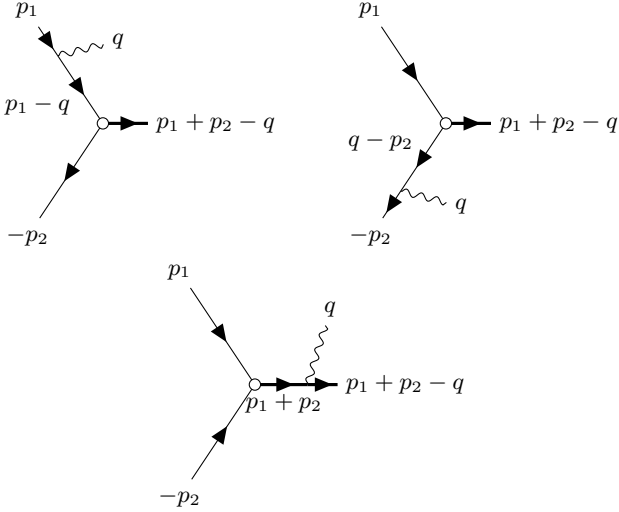
\begin{figure}
  \centering
  \begin{tikzpicture}
    \begin{feynman}
      \vertex[empty dot] (m) at ( 0, 0) {}; 
      \vertex (a) at (-1.,1.5) {$p_1$};
      \vertex (b) at (-1.,-1.5) {$-p_2$};
      \vertex (c) at (1.5, 0) {$p_1+p_2-q$};
      \vertex (v) at ( -0.6,0.9);
      \vertex (e) at (0.2,1.1) {$q$};
      \diagram* {
        (a) -- [fermion] (v) -- [fermion,edge label'=$p_1-q$] (m) -- [fermion, very thick] (c),
        (m) -- [fermion] (b), 
        (v) -- [photon] (e),
       };
    \end{feynman}
  \end{tikzpicture}
\quad
\quad
  \begin{tikzpicture}
    \begin{feynman}
      \vertex[empty dot] (m) at ( 0, 0) {}; 
      \vertex (a) at (-1.,1.5) {$p_1$};
      \vertex (b) at (-1.,-1.5) {$-p_2$};
      \vertex (c) at (1.5, 0) {$p_1+p_2-q$};
      \vertex (v) at ( -0.6,-0.9);
      \vertex (e) at (0.2,-1.1) {$q$};
      \diagram* {
        (a) -- [fermion] (m) -- [fermion, very thick] (c),
        (m) -- [fermion,edge label'=$q-p_2$] (v) -- [fermion] (b), 
        (v) -- [photon] (e),
       };
    \end{feynman}
  \end{tikzpicture}

  \begin{tikzpicture}
    \begin{feynman}
      \vertex[empty dot] (m) at ( 0, 0) {}; 
      \vertex (a) at (-1.,1.5) {$p_1$};
      \vertex (b) at (-1.,-1.5) {$-p_2$};
      \vertex (c) at (2., 0) {$p_1+p_2-q$};
      \vertex (v) at (0.7, 0);
      \vertex (e) at (1., 1.) {$q$};
      \diagram* {
        (a) -- [fermion] (m) -- [fermion,edge label'=$p_1+p_2$, very thick] (v) -- [fermion, very thick] (c),
        (b) -- [fermion] (m) ,
        (v) -- [photon] (e),
      };
    \end{feynman}
  \end{tikzpicture}

  \caption{Bremsstrahlung photon from quark and antiquark fussion to meson, ignoring possible emission and/or absorption of soft gluons}
\label{meson-fig}
\end{figure}

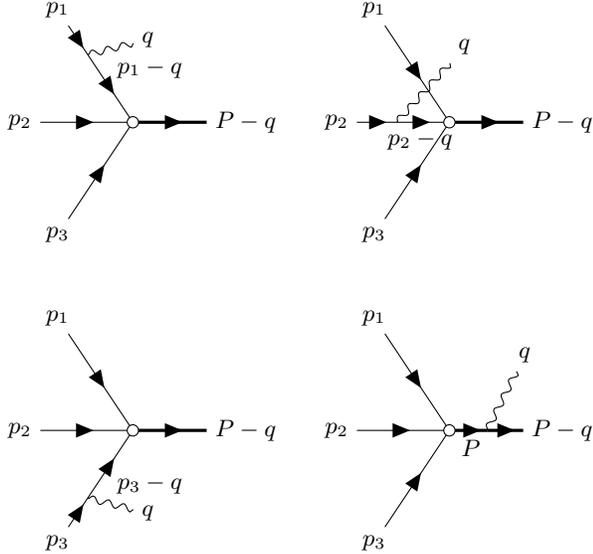
\begin{figure}
\centerline{
  \begin{tikzpicture}
    \begin{feynman}
      \vertex[empty dot] (m) at ( 0, 0) {}; 
      \vertex (a) at (-1.,1.5) {$p_1$};
      \vertex (b) at (-1.,-1.5) {$p_3$};
      \vertex (c) at (1.5, 0) {$P-q$};
      \vertex (d) at (-1.5, 0) {$p_2$};
      \vertex (v) at ( -0.6,0.9);
      \vertex (e) at (0.2,1.1) {$q$};
      \diagram* {
        (a) -- [fermion] (v) -- [fermion,edge label=$p_1-q$] (m) -- [fermion, very thick] (c),
        (b) -- [fermion] (m), 
        (d) -- [fermion] (m), 
        (v) -- [photon] (e),
       };
    \end{feynman}
  \end{tikzpicture}
\quad
  \begin{tikzpicture}
    \begin{feynman}
      \vertex[empty dot] (m) at ( 0, 0) {}; 
      \vertex (a) at (-1.,1.5) {$p_1$};
      \vertex (b) at (-1.,-1.5) {$p_3$};
      \vertex (c) at (1.5, 0) {$P-q$};
      \vertex (d) at (-1.5, 0) {$p_2$};
      \vertex (v) at (-0.7, 0);
      \vertex (e) at (0.2, 1.) {$q$};
      \diagram* {
        (a) -- [fermion] (m) -- [fermion, very thick] (c),
        (d) -- [fermion] (v) -- [fermion,edge label'=$p_2-q$] (m), 
        (b) -- [fermion] (m), 
        (v) -- [photon] (e),
       };
    \end{feynman}
  \end{tikzpicture}
}
\vspace{2em}
\centerline{
  \begin{tikzpicture}
    \begin{feynman}
      \vertex[empty dot] (m) at ( 0, 0) {}; 
      \vertex (a) at (-1.,1.5) {$p_1$};
      \vertex (b) at (-1.,-1.5) {$p_3$};
      \vertex (c) at (1.5, 0) {$P-q$};
      \vertex (d) at (-1.5, 0) {$p_2$};
      \vertex (v) at ( -0.6,-0.9);
      \vertex (e) at (0.2,-1.1) {$q$};
      \diagram* {
        (a) -- [fermion] (m) -- [fermion, very thick] (c),
        (b) -- [fermion] (v) -- [fermion,edge label'=$p_3-q$] (m), 
        (d) -- [fermion] (m), 
        (v) -- [photon] (e),
       };
    \end{feynman}
  \end{tikzpicture}
\quad
  \begin{tikzpicture}
    \begin{feynman}
      \vertex[empty dot] (m) at ( 0, 0) {}; 
      \vertex (a) at (-1.,1.5) {$p_1$};
      \vertex (b) at (-1.,-1.5) {$p_3$};
      \vertex (c) at (1.5, 0) {$P-q$};
      \vertex (d) at (-1.5, 0) {$p_2$};
      \vertex (v) at (0.5, 0);
      \vertex (e) at (1., 1.) {$q$};
      \diagram* {
        (a) -- [fermion] (m), 
        (d) -- [fermion] (m) -- [fermion,edge label'=$P$, very thick] (v) -- [fermion, very thick] (c),
        (b) -- [fermion] (m), 
        (v) -- [photon] (e),
       };
    \end{feynman}
  \end{tikzpicture}
}
  \caption{Bremsstrahlung photon from three quark fussion to baryon where $P=p_1+p_2+p_3$, ignoring possible emission and/or absorption of soft gluons}
\label{baryon-fig}
\end{figure}

The momentum distribution of radiated photon from decelerated charged particle $i$ is expressed as~\cite{Low:1958sn,Koch:1990jzd,Peskin:1995ev}
\begin{eqnarray}
\omega\frac{dN^\gamma}{d^3 {\bf q}}=\frac{1}{2(2\pi)^3}\sum_\lambda| j_i\cdot\varepsilon^\lambda(q)|^2,
\label{spec-eq}
\end{eqnarray}
where $q_\mu=(\omega,~{\bf q})$ is photon energy and momentum, $(j_i)_\mu$ electromagnetic current induced by the charged particle $i$, and $\varepsilon_\mu^\lambda(q)$ the polarization vector of emitted photon with $\lambda$ being the polarization state.

In Fig.~\ref{meson-fig} the transition amplitude for quark and antiquark to form a meson with a photon emission is proportional to~\cite{Low:1958sn,Koch:1990jzd,Peskin:1995ev}
\begin{eqnarray}
J\cdot \varepsilon^\lambda=\bigg\{-Q_1\frac{p_1^\mu}{p_1\cdot q}-Q_2\frac{p_2^\mu}{p_2\cdot q}\nonumber\\
+Q_m\frac{(p_1+p_2-q)^\mu}{(p_1+p_2-q)\cdot q}\bigg\}\varepsilon_\mu^\lambda (q),
\label{meson}
\end{eqnarray}
where $Q_1$ and $Q_2$ are, respectively, the electric charges of quark and antiquark and $Q_m=Q_1+Q_2$.

As for the three quarks which form a baryon,
\begin{eqnarray}
J\cdot \varepsilon^\lambda=\bigg\{-Q_1\frac{p_1^\mu}{p_1\cdot q}-Q_2\frac{p_2^\mu}{p_2\cdot q}-Q_3\frac{p_3^\mu}{p_3\cdot q}\nonumber\\
+Q_b\frac{(p_1+p_2+p_3-q)^\mu}{(p_1+p_2+p_3-q)\cdot q}\bigg\}\varepsilon_\mu^\lambda (q),
\label{baryon}
\end{eqnarray}
as shown in Fig.~\ref{baryon-fig}, with $Q_1$, $Q_2$ and $Q_3$ being electric charges of three quarks and $Q_b=Q_1+Q_2+Q_3$.

Substituting Eqs.~(\ref{meson}) and (\ref{baryon}) into Eq.~(\ref{spec-eq}) and using $\sum_\lambda \varepsilon_\mu^{\lambda *} (q)\varepsilon_\nu^\lambda (q)=-g_{\mu\nu}$, the photon spectrum from meson formation is given by
\begin{eqnarray}
\omega\frac{dN_m^\gamma}{d^3 {\bf q}}=\frac{1}{2(2\pi)^3}\bigg[-Q_1^2\frac{m_1^2}{(p_1\cdot q)^2}-Q_2^2\frac{m_2^2}{(p_2\cdot q)^2}\nonumber\\
-Q_m^2\frac{(P-q)^2}{((P-q)\cdot q)^2}-2Q_1 Q_2\frac{p_1\cdot p_2}{(p_1\cdot q)(p_2\cdot q)}\nonumber\\
+\frac{2Q_m}{(P-q)\cdot q}\bigg\{Q_1\frac{p_1\cdot (P-q)}{(p_1\cdot q)}+Q_2\frac{p_2\cdot (P-q)}{(p_2\cdot q)}\bigg\}\bigg],
\label{meson2}
\end{eqnarray}
where $P=p_1+p_2$, and that from baryon formation
\begin{eqnarray}
\omega\frac{dN_b^\gamma}{d^3 {\bf q}}=\frac{1}{2(2\pi)^3}\bigg[-\sum_{i,j=1,2,3}Q_iQ_j\frac{p_i p_j}{(p_i\cdot q)(p_j\cdot q)}\nonumber\\
-Q_b^2\frac{(P-q)^2}{((P-q)\cdot q)^2}\nonumber\\
+\frac{2Q_b}{(P-q)\cdot q}\sum_{i=1,2,3}Q_i\frac{p_i\cdot (P-q)}{(p_i\cdot q)}\bigg],
\label{baryon2}
\end{eqnarray}
where $P=p_1+p_2+p_3$.

The above calculations are called the soft photon approximation, because each term in the curly brackets of Eqs.~(\ref{meson}) and (\ref{baryon}) is a propagator approximated in the limit of soft photon ($q \ll p_i$), and it is assumed that the photon emission does not affect the main scattering which causes charge deceleration~\cite{Song:2022wil}. 
For the approximation to be valid, photon energy should be much smaller than scattering energy.

Figs.~\ref{meson-fig} and \ref{baryon-fig} are reminiscent of the coalescence model which is widely used to describe the hadronization of quarks and antiquarks~\cite{Greco:2003xt}.
One drawback of the model is that total energy is not conserved, for example, the pion formation from a quark and antiquark pair.
More correct or realistic Feynman diagrams will be accompanied by soft gluon emission and/or absorption, which affects the hadronization of other partons~\cite{Song:2021mvc}.
For instance, if the invariant mass of combined partons is smaller than the hadron mass, the deficient energy is supplied by absorbing soft gluons, while soft gluons are emitted in the opposite case.
The gluon emission or absorption is more favored than that of photon because of much larger strong coupling than the coupling in QED.
Therefore, the hadronization in Figs.~\ref{meson-fig} and \ref{baryon-fig} is not simply $n\rightarrow 1$, but it involves soft gluon emission and/or absorption, and the scattering energy is different from hadron mass.
In the present study a photon is then attached to the hadronization process following the soft photon approximation.

\section{photon from a hadroning QGP}\label{QGP}

For simplicity, we assume that the transition amplitudes for (anti)quarks to form meson ($A_m$), baryon ($A_b$) and anti-baryon ($A_{\bar{b}}$) at $T_c$ are constants as follows:
\begin{eqnarray}
&&A_m \sum_{i,j=u,d,s}\int \frac{d^3p_1}{(2\pi)^3}\frac{d^3p_2}{(2\pi)^3} f_i(p_1)f_{\bar{j}}(p_2)=n_m, \label{am}\\
&&A_b\sum_{i,j,k=u,d,s}\int \frac{d^3p_1}{(2\pi)^3}\frac{d^3p_2}{(2\pi)^3}\frac{d^3p_3}{(2\pi)^3}\nonumber\\ 
&&~~~~~~~~~~~~~~~~~~\times f_i(p_1)f_j(p_2)f_k(p_3)=n_b, \label{ab}\\
&&A_{\bar{b}}\sum_{i,j,k=u,d,s}\int \frac{d^3p_1}{(2\pi)^3}\frac{d^3p_2}{(2\pi)^3}\frac{d^3p_3}{(2\pi)^3}\nonumber\\ 
&&~~~~~~~~~~~~~~~~~~\times f_{\bar{i}}(p_1)f_{\bar{j}}(p_2)f_{\bar{k}}(p_3)=n_{\bar{b}}
\label{coeff}
\end{eqnarray}
where $f_i(p)$ is the Fermi-Dirac distribution of the parton $i$, including spin-color degeneracy, and $n_m$, $n_b$ and $n_{\bar{b}}$ are respectively the number densities of meson, baryon and anti-baryon at $T_c=$ 0.16 GeV.
In the statistical model 
$n_m=0.35 ~{\rm fm^{-3}}$ and $n_b=n_{\bar{b}}=0.032~ {\rm fm^{-3}}$
for $\mu_B=$ 0, and $n_m=0.35 ~{\rm fm^{-3}}$, $n_b=0.056~ {\rm fm^{-3}}$ and $n_{\bar{b}}=0.0046~ {\rm fm^{-3}}$ for $\mu_B=$ 200 MeV.
The former $\mu_B$ corresponds to LHC and RHIC energies and the latter $\mu_B$ to SPS energy~\cite{Andronic:2021dkw}.

$f_i(p)$ depends on the effective mass of (anti)quark at $T_c$, which can be obtained from the assumption that all (anti)quarks are consumed through hadronization, that is, 
\begin{eqnarray}
\sum_{i=u,d,s}\int \frac{d^3p}{(2\pi)^3} f_i(p)=n_m+3n_b, \nonumber\\
\sum_{i=u,d,s}\int \frac{d^3p}{(2\pi)^3} f_{\bar{i}}(p)=n_m+3n_{\bar{b}}.
\end{eqnarray} 

Assuming that strange quark mass is same as up/down quark mass for simplicity, (anti)quark mass is about 340 MeV at $\mu_B=$ 0, and from Eq.~(\ref{coeff})
\begin{eqnarray}
A_m&=&n_m/(N_f n_q)^2=2.27 \times 10^2~{\rm GeV^{-3}}, \nonumber\\
A_b&=&A_{\bar{b}}=n_b/(N_f n_q)^3=3.02 \times 10^3~{\rm GeV^{-6}}
\label{mu0}
\end{eqnarray} 
with flavor number $N_f=3$ and quark number density $n_q=\int d^3p/(2\pi)^3 f_q(p)$.

For $\mu_B=$ 200 MeV, (anti)quark mass is taken to be 310 MeV and 
\begin{eqnarray}
A_m&=&n_m/(N_f n_q)/(N_f n_{\bar{q}})=1.87 \times 10^2~{\rm GeV^{-3}}, \nonumber\\
A_b&=&n_b/(N_f n_q)^3=2.96 \times 10^3~{\rm GeV^{-6}},\nonumber\\
A_{\bar{b}}&=&n_{\bar{b}}/(N_f n_{\bar{q}})^3=1.72 \times 10^3~{\rm GeV^{-6}}.
\label{mu200}
\end{eqnarray} 
We note that in Eq.~(\ref{mu200}) the quark chemical potential is 54 MeV, which is not exactly one third $\mu_B$.

Photon production from unit volume of hadronized QGP 
is then given by 
\begin{eqnarray}
\omega\frac{dN^\gamma}{V d^3 {\bf q}}=A_m\sum_{i,j=u,d,s}\int \frac{d^3p_1}{(2\pi)^3}\frac{d^3p_2}{(2\pi)^3}~~~~~~~~~~\nonumber\\ \times f_i(p_1)f_{\bar{j}}(p_2)\omega\frac{dN_m^\gamma}{d^3 {\bf q}}\nonumber\\
+A_b\sum_{i,j,k=u,d,s}\int \frac{d^3p_1}{(2\pi)^3}\frac{d^3p_2}{(2\pi)^3}\frac{d^3p_3}{(2\pi)^3}~~~\nonumber\\\times f_i(p_1)f_j(p_2)f_k(p_3)\omega\frac{dN_b^\gamma}{d^3 {\bf q}}\nonumber\\
+A_{\bar{b}}\sum_{i,j,k=u,d,s}\int \frac{d^3p_1}{(2\pi)^3}\frac{d^3p_2}{(2\pi)^3}\frac{d^3p_3}{(2\pi)^3}~~~\nonumber\\ \times f_{\bar{i}}(p_1)f_{\bar{j}}(p_2)f_{\bar{k}}(p_3)\omega\frac{dN_{\bar{b}}^\gamma}{d^3 {\bf q}}.
\end{eqnarray}

The upper limit of photon energy from meson formation is given by
\begin{eqnarray}
(p_1+p_2-q)^2\ge m_\pi^2,
\label{conserv1}
\end{eqnarray}
and for (anti)baryon formation by
\begin{eqnarray}
(p_1+p_2+p_3-q)^2\ge m_N^2
\label{conserv2}
\end{eqnarray}
with $m_\pi$ and $m_N$ being respectively pion and nucleon masses.

\begin{figure} [h]
\centerline{
\includegraphics[width=8.6 cm]{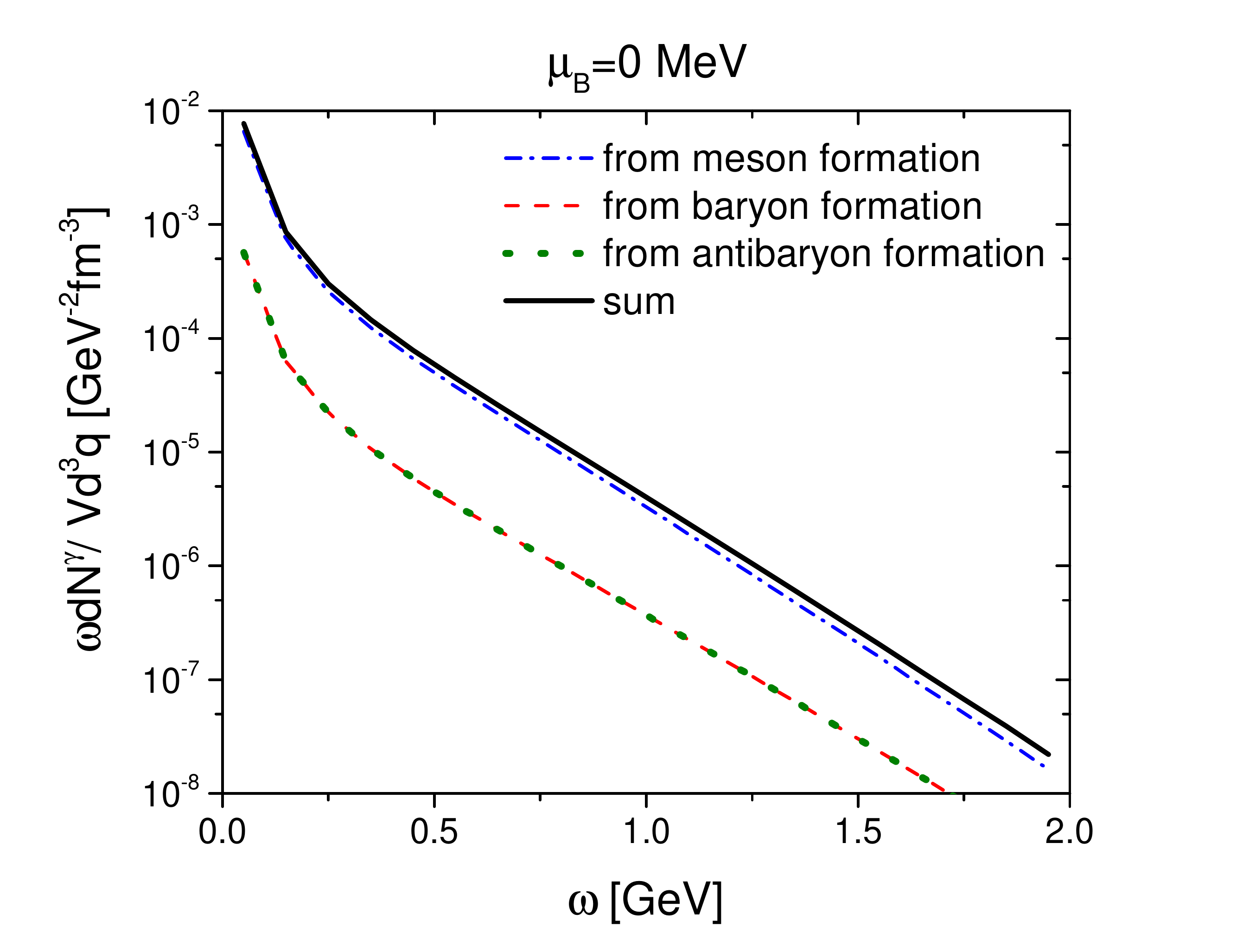}}
\centerline{
\includegraphics[width=8.6 cm]{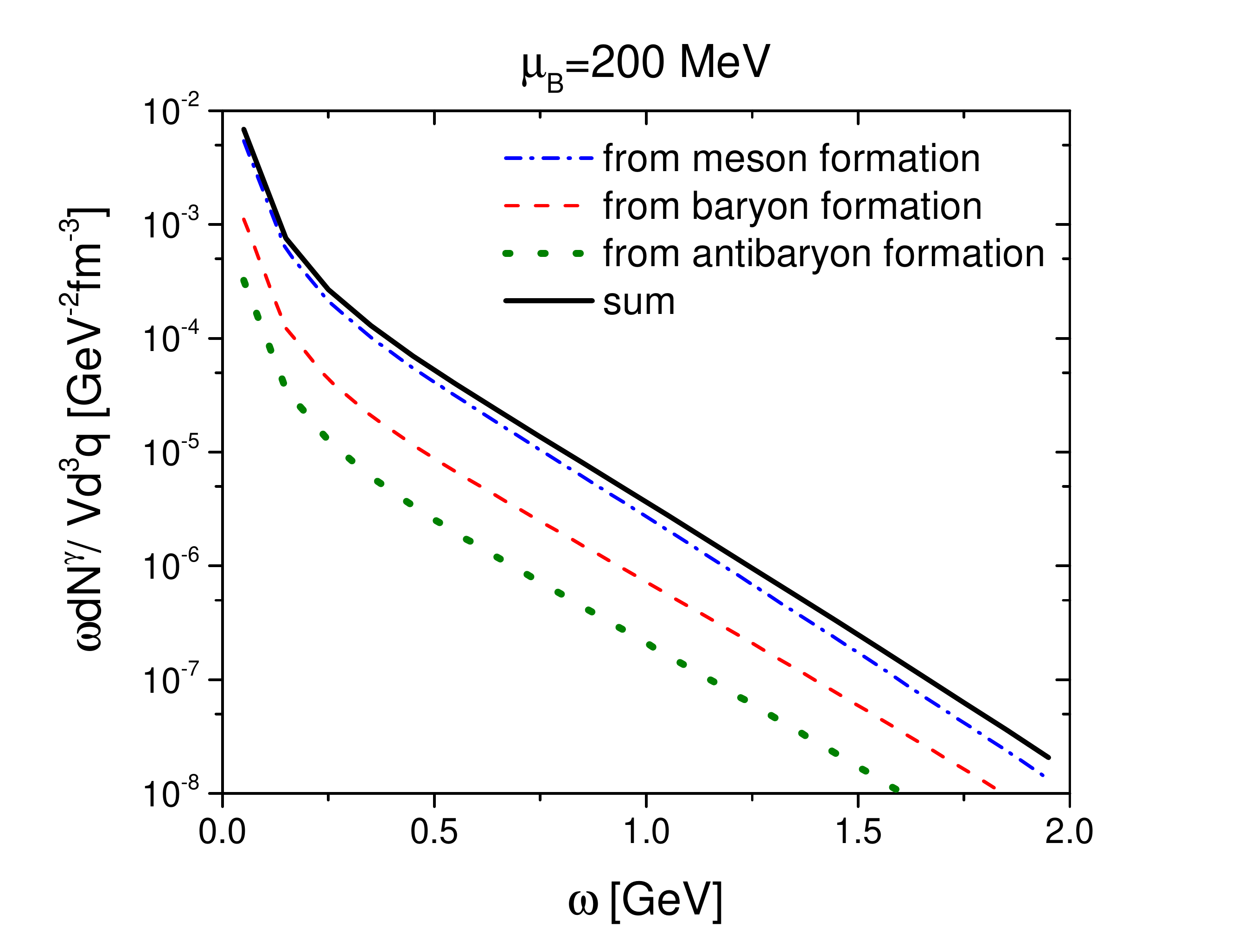}}
\caption{(Color online) Bremsstrahlung photon spectrum from unit volume (1 $\rm fm^3$) of hadronizing QGP at $T_c=$ 160 MeV
for $\mu_B=0$ (upper) and for $\mu_B=200$ MeV (lower)} \label{spec}
\end{figure}

The photon energy should be much smaller than the scattering energy for the soft photon approximation to be valid~\cite{Eggers:1995jq}. 
In thermal equilibrium the parton distribution function is peaked around temperature and so is the scattering energy.
However, it does not mean that all thermal partons have little momentum at $T_{c}$, because some of them still have large momentum, as shown in the spectrum of thermal photon from QGP or from hadron gas (HG), which does not terminate at low $p_T$ but reaches high $p_T$~\cite{Paquet:2015lta}.

Though the photon energy is always smaller than the scattering energy according to the energy conservation in Eqs.~(\ref{conserv1}) and (\ref{conserv2}), the condition for the soft photon approximation $(\omega\ll \sqrt{s})$
is not well satisfied as photon energy increases.
The scattering cross section expanded in term of photon energy divided by the energy scale of the scattering, only the leading term is taken in this study. 
There are systematic studies on the second leading terms in the expansion~\cite{Low:1958sn,Burnett:1967km}, which require the derivative of transition amplitue with respect to external momentum.
Since we assume constant transition amplitudes for hadronization in Eqs.~(\ref{am}), (\ref{ab}) and (\ref{coeff}), it is not possible to calculate the second leading terms in the present form. 
If the photon energy is not much smaller than the energy scale, subleading corrections 
will not be negligable~\cite{Low:1958sn,Eggers:1995jq}.
Since the underlying mechanism for hadronization is not
well understood and/or not yet known, it is presently unclear whether the subleading terms will enhance or suppress the Bremsstrahlung photon.

Figure~\ref{spec} shows the spectra of Bremsstrahlung photon from unit volume (1 $\rm fm^3$) of hadronizing QGP at $T_c=$ 160 MeV for $\mu_B=0$ and $\mu_B=200$ MeV.
The contribution from meson formation is larger than that from (anti)baryon formation, since the meson density is much larger than that of (anti)baryon.
On the other hand, the Bremsstrahlung photon from (anti)baryon formation is a bit harder than that from meson formation.
We note that the spectrum for $\mu_B=0$ is about 10 \% larger at low momentum than that for $\mu_B=200$ MeV partly due to the differences between Eqs.~(\ref{mu0}) and (\ref{mu200}).
It is interesting to see that the photon spectrum in Fig.~\ref{spec} is comparable to the photon production rate in QGP as well as in HG near $T_c$, which is shown in Fig. 3 of Ref.~\cite{Paquet:2015lta}.
We note that the coalescence photons are produced along a hypersurface at $T_c$ while the the production rate is thermal photons produced per unit time (1 fm/c) in matter.

\begin{figure} [h]
\centerline{
\includegraphics[width=8.6 cm]{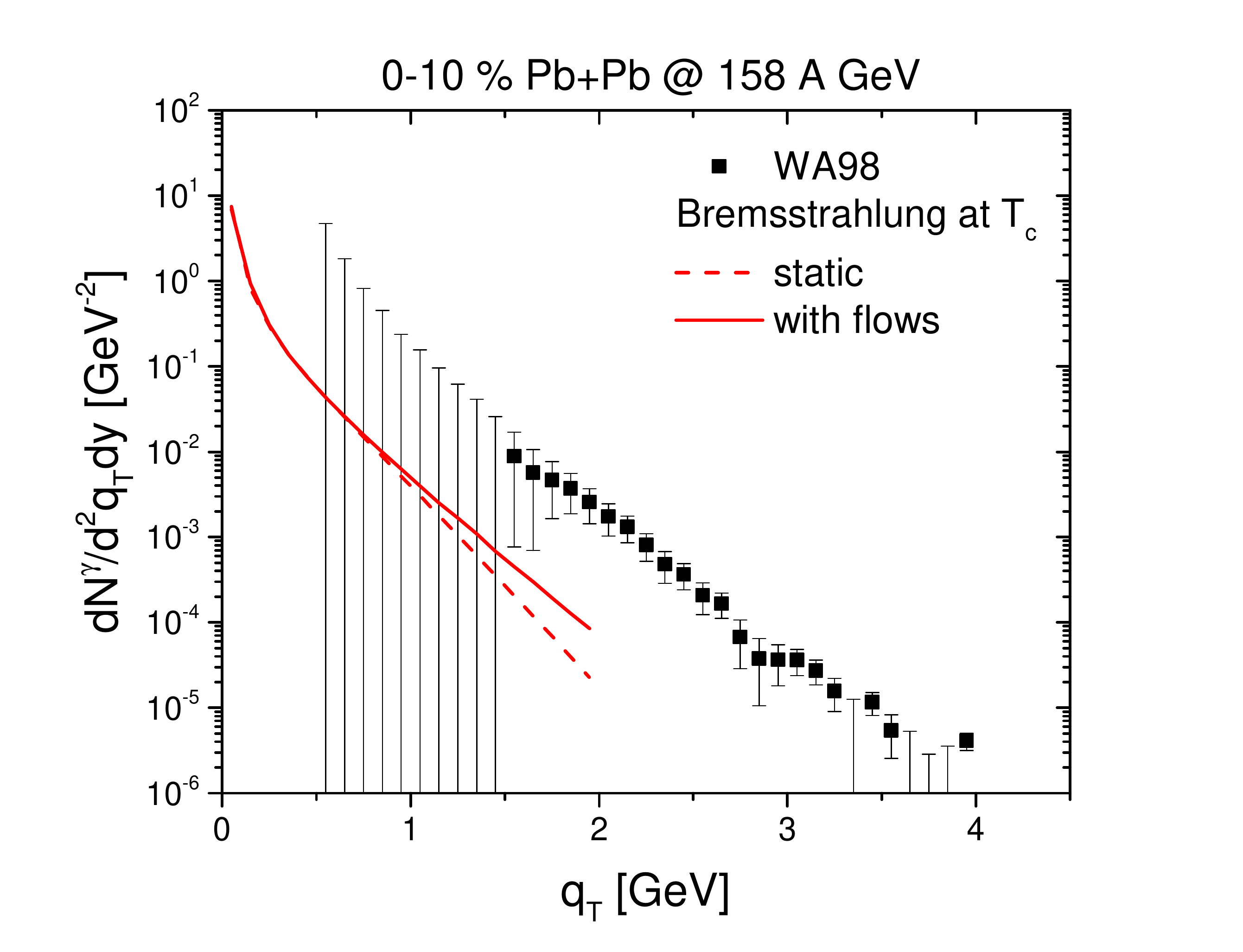}}
\centerline{
\includegraphics[width=8.6 cm]{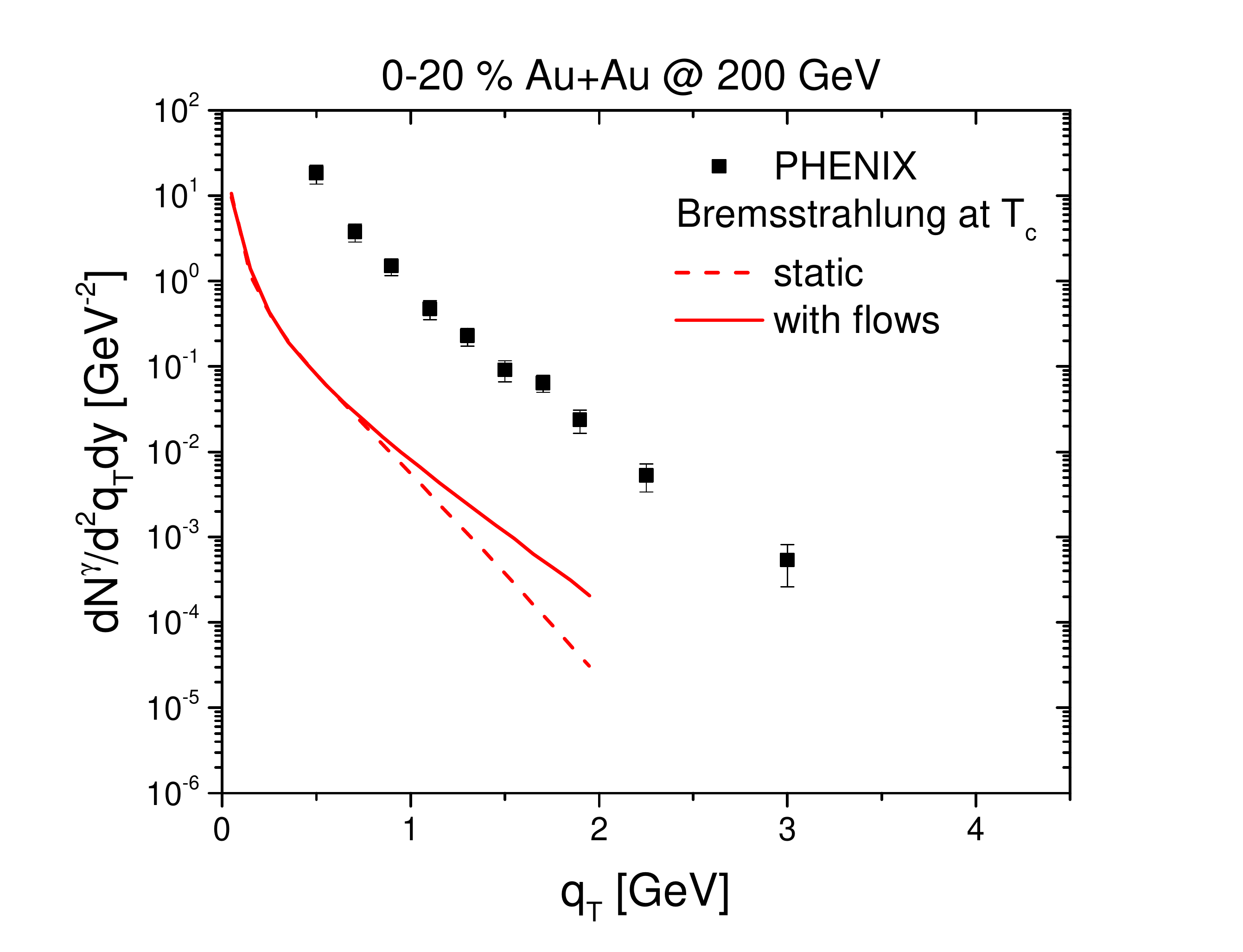}}
\caption{(Color online) Bremsstrahlung photon spectra as a function of transverse momentum in 0-10 \% central Pb+Pb collisions at $E_{\rm kin.}=$ 158 A GeV and in 0-20 \% central Au+Au collisions at $\sqrt{s_{\rm NN}}=$ 200 GeV with and without considering transverse flows, compared with experimental data from the WA98 and PHENIX Collaborations~\cite{WA98:2000ulw,PHENIX:2014nkk}} \label{exp-pt}
\end{figure}

One can compare the results with the experimental data in heavy-ion collisions, taking into account the volume of QGP at $T_c$, which
can be deduced from charged particle or pion yield, assuming that the yield does not change after the chemical freeze-out temperature which is almost same as $T_c$~\cite{Andronic:2021dkw}.
In the statistical model the number density of charged particles at $T=$ 160 MeV is 0.367 ${\rm fm^{-3}}$, and that of charged pions 0.325 ${\rm fm^{-3}}$ including the feed-down from decays. 
The charged particle yield at mid-rapidity in 0-10 \% central Pb+Pb collisions at 158 A GeV is roughly 400~\cite{NA50:2002edr}, which can be interpreted to 1090 ${\rm fm^3}$ of QGP volume at $T_c$. Similarly the charged pion yield in 0-20 \% central Au+Au collisions at $\sqrt{s_{\rm NN}}=$ 200 GeV, which is 448, is interpreted to 1380 ${\rm fm^3}$.

Figure~\ref{exp-pt} compares our results with the experimental data from the WA98 and PHENIX Collaborations~\cite{WA98:2000ulw,PHENIX:2014nkk}.  
The dashed lines are the spectra without considering transverse flows and the solid lines the spectra with including the transverse flows whose velocities are, respectively, 0.33 and 0.44 at $T_c$ in the upper and lower panels from a schematic hydrodynamics~\cite{Song:2010fk}.
Comparing the spectra in the upper and lower panels, one can see that the contribution from the Bremsstrahlung photon at $T_c$ is larger in central Pb+Pb collisions at $E_{\rm kin.}=$ 158 A GeV than in central Au+Au collisions at $\sqrt{s_{\rm NN}}=$ 200 GeV.
The reason is that the direct photon is continually produced from or even before the formation of QGP to the freeze-out in heavy-ion collisions, while the Bremsstrahlung photon from a hadronizing QGP is produced only once at $T_c$.
Though the strong coupling $\alpha_s$ is large near $T_c$ to force all partons to hadronize, 
 abundant thermal photons overshine the Bremsstrahlung photons at $T_c$, if the lifetime of QGP is long as in the central Au+Au collisions at $\sqrt{s_{\rm NN}}=$ 200 GeV. 
It means that the Bremsstrahlung photon from a hadronzing QGP will be more visible in lower-energy heavy-ion collisions, and maximized when the collision energy of heavy-ions barely reaches the phase boundary.

\begin{figure} [h]
\centerline{
\includegraphics[width=8.6 cm]{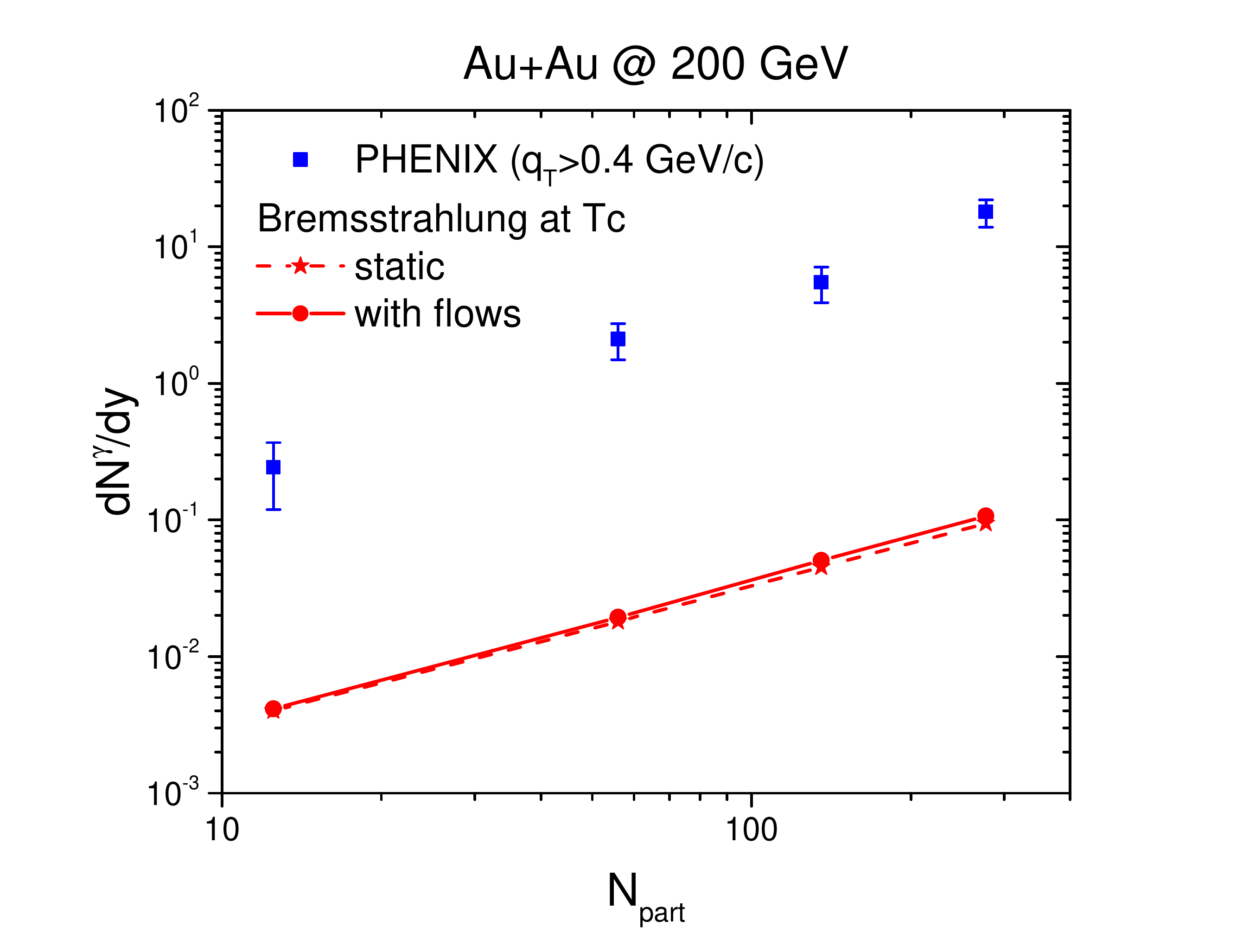}}
\caption{(Color online) Bremsstrahlung photon yield per rapidity for $q_T>$ 0.4 GeV in Au+Au collisions at $\sqrt{s_{\rm NN}}=$ 200 GeV as a function of the number of participants in comparison with the experimental data from the PHENIX Collaboration~\cite{PHENIX:2014nkk}} \label{rhic}
\end{figure}

Figure~\ref{rhic} displays the Bremsstrahlung photon yield per rapidity for $q_T>$ 0.4 GeV in Au+Au collisions at $\sqrt{s_{\rm NN}}=$ 200 GeV as a function of the number of participants, compared with the experimental data from the PHENIX Collaboration~\cite{PHENIX:2014nkk}.
Pion yields in 20-40, 40-60 and 60-92 \% centralities from the PHENIX Collaboration~\cite{PHENIX:2003iij} are, respectively, converted to 660, 264 and 58 ${\rm fm^3}$ of QGP volume at $T_c$.
The dashed line does not include flow effects and the solid line includes transverse flows at $T_c$, whose velocities are respectively 0.41, 0.33 and 0.21 in 20-40, 40-60 and 60-92 \% centralities from the schematic hydrodynamics~\cite{Song:2010fk}. 
Fig.~\ref{rhic} shows that the contribution from the Bremsstrahlung photon at $T_c$ increases from central collisions to peripheral collisions. 
The reason is that the lifetime of nuclear matter is relatively short in peripheral collisions, which is consistent with Fig.~\ref{exp-pt}.

\section{summary}\label{summary}

The direct photon excluding the decay photon and the prompt photon is produced from a nuclear matter in heavy-ion collisions.
There have been many studies on the photon production from QGP and from hadron gas, but the photon production from hadronization has barely been studied.
In this work we have estimated, by using the soft photon approximation, the production of Bremsstrahlung photon at hadronization where charged quarks and antiquarks change their momenta.
For the calculations it is assumed that quark and antiquark numbers do not change during the hadronization and that a quark and an antiquark form a meson, three quarks a baryon and three antiquarks an antibaryon.
Since the statistical model provides the number density of each species of hadrons near $T_c$, one can deduce the number densities of quarks and antiquarks at $T_c$.
For this, (anti)quark mass is taken to be 340 MeV for $\mu_B=$ 0 and 310 MeV for $\mu_B=$ 200 MeV, which are consistent with the constituent quark mass.
The constant transition amplitudes for meson, baryon and antibaryon formations are obtained from that all quarks and antiquarks must be consumed in hadronization.

We have found that the Bremsstrahlung photon from meson formation is dominant over that from (anti)baryon formation, because mesons are more produced than (anti)baryons at $T_c$ in the statistical model.
But the spectrum from baryon formation is a bit harder than that of meson formation.
We have also found that Bremsstrahlung photon yield from hadronization is about 10 \% less at $\mu_B=$ 200 MeV than at $\mu_B=$ 0.
It is interesting that the Bremsstrahlung photon from a hadronizing QGP is comparable to the production rate of thermal photon per unit time (1 fm/c) in QGP or in HG near $T_c$.

The number of charged particles or of charged pions interpreted to the volume of QGP, our calculations are compared with the experimental data on direct photon in heavy-ion collisions.
We have found that the contribution from the Bremsstrahlung photon at $T_c$ to the direct photon produced in heavy-ion collisions increases in low-energy collisions and in peripheral collisions, because the thermal photon is continually produced from the initial stage of a hot dense nuclear matter to the freeze-out and its yield is proportional to the lifetime of the matter.
Therefore, the contribution from the Bremsstrahlung photon at $T_c$ will be maximized, when the collision energy of heavy-ions barely reaches the phase boundary, and it may serve as a signal for the QGP formation in heavy-ion collisions.

\section*{Acknowledgements}
The computational resources have been provided by the LOEWE-Center for
Scientific Computing and the "Green Cube" at GSI, Darmstadt.

\bibliography{revision2}

\begin{thebibliography}{33}
\expandafter\ifx\csname natexlab\endcsname\relax\def\natexlab#1{#1}\fi
\expandafter\ifx\csname bibnamefont\endcsname\relax
  \def\bibnamefont#1{#1}\fi
\expandafter\ifx\csname bibfnamefont\endcsname\relax
  \def\bibfnamefont#1{#1}\fi
\expandafter\ifx\csname citenamefont\endcsname\relax
  \def\citenamefont#1{#1}\fi
\expandafter\ifx\csname url\endcsname\relax
  \def\url#1{\texttt{#1}}\fi
\expandafter\ifx\csname urlprefix\endcsname\relax\def\urlprefix{URL }\fi
\providecommand{\bibinfo}[2]{#2}
\providecommand{\eprint}[2][]{\url{#2}}

\bibitem[{\citenamefont{Linnyk et~al.}(2016)\citenamefont{Linnyk, Bratkovskaya,
  and Cassing}}]{Linnyk:2015rco}
\bibinfo{author}{\bibfnamefont{O.}~\bibnamefont{Linnyk}},
  \bibinfo{author}{\bibfnamefont{E.~L.} \bibnamefont{Bratkovskaya}},
  \bibnamefont{and} \bibinfo{author}{\bibfnamefont{W.}~\bibnamefont{Cassing}},
  \bibinfo{journal}{Prog. Part. Nucl. Phys.} \textbf{\bibinfo{volume}{87}},
  \bibinfo{pages}{50} (\bibinfo{year}{2016}), \eprint{1512.08126}.

\bibitem[{\citenamefont{David}(2020)}]{David:2019wpt}
\bibinfo{author}{\bibfnamefont{G.}~\bibnamefont{David}},
  \bibinfo{journal}{Rept. Prog. Phys.} \textbf{\bibinfo{volume}{83}},
  \bibinfo{pages}{046301} (\bibinfo{year}{2020}), \eprint{1907.08893}.

\bibitem[{\citenamefont{Shen et~al.}(2014)\citenamefont{Shen, Heinz, Paquet,
  and Gale}}]{Shen:2013vja}
\bibinfo{author}{\bibfnamefont{C.}~\bibnamefont{Shen}},
  \bibinfo{author}{\bibfnamefont{U.~W.} \bibnamefont{Heinz}},
  \bibinfo{author}{\bibfnamefont{J.-F.} \bibnamefont{Paquet}},
  \bibnamefont{and} \bibinfo{author}{\bibfnamefont{C.}~\bibnamefont{Gale}},
  \bibinfo{journal}{Phys. Rev. C} \textbf{\bibinfo{volume}{89}},
  \bibinfo{pages}{044910} (\bibinfo{year}{2014}), \eprint{1308.2440}.

\bibitem[{\citenamefont{Paquet et~al.}(2016)\citenamefont{Paquet, Shen,
  Denicol, Luzum, Schenke, Jeon, and Gale}}]{Paquet:2015lta}
\bibinfo{author}{\bibfnamefont{J.-F.} \bibnamefont{Paquet}},
  \bibinfo{author}{\bibfnamefont{C.}~\bibnamefont{Shen}},
  \bibinfo{author}{\bibfnamefont{G.~S.} \bibnamefont{Denicol}},
  \bibinfo{author}{\bibfnamefont{M.}~\bibnamefont{Luzum}},
  \bibinfo{author}{\bibfnamefont{B.}~\bibnamefont{Schenke}},
  \bibinfo{author}{\bibfnamefont{S.}~\bibnamefont{Jeon}}, \bibnamefont{and}
  \bibinfo{author}{\bibfnamefont{C.}~\bibnamefont{Gale}},
  \bibinfo{journal}{Phys. Rev. C} \textbf{\bibinfo{volume}{93}},
  \bibinfo{pages}{044906} (\bibinfo{year}{2016}), \eprint{1509.06738}.

\bibitem[{\citenamefont{Adare et~al.}(2015)}]{PHENIX:2014nkk}
\bibinfo{author}{\bibfnamefont{A.}~\bibnamefont{Adare}} \bibnamefont{et~al.}
  (\bibinfo{collaboration}{PHENIX}), \bibinfo{journal}{Phys. Rev. C}
  \textbf{\bibinfo{volume}{91}}, \bibinfo{pages}{064904}
  (\bibinfo{year}{2015}), \eprint{1405.3940}.

\bibitem[{\citenamefont{Adare et~al.}(2016)}]{PHENIX:2015igl}
\bibinfo{author}{\bibfnamefont{A.}~\bibnamefont{Adare}} \bibnamefont{et~al.}
  (\bibinfo{collaboration}{PHENIX}), \bibinfo{journal}{Phys. Rev. C}
  \textbf{\bibinfo{volume}{94}}, \bibinfo{pages}{064901}
  (\bibinfo{year}{2016}), \eprint{1509.07758}.

\bibitem[{\citenamefont{Adam et~al.}(2016)}]{ALICE:2015xmh}
\bibinfo{author}{\bibfnamefont{J.}~\bibnamefont{Adam}} \bibnamefont{et~al.}
  (\bibinfo{collaboration}{ALICE}), \bibinfo{journal}{Phys. Lett. B}
  \textbf{\bibinfo{volume}{754}}, \bibinfo{pages}{235} (\bibinfo{year}{2016}),
  \eprint{1509.07324}.

\bibitem[{\citenamefont{Acharya et~al.}(2019)}]{ALICE:2018dti}
\bibinfo{author}{\bibfnamefont{S.}~\bibnamefont{Acharya}} \bibnamefont{et~al.}
  (\bibinfo{collaboration}{ALICE}), \bibinfo{journal}{Phys. Lett. B}
  \textbf{\bibinfo{volume}{789}}, \bibinfo{pages}{308} (\bibinfo{year}{2019}),
  \eprint{1805.04403}.

\bibitem[{\citenamefont{Eggers et~al.}(1996)\citenamefont{Eggers, Tabti, Gale,
  and Haglin}}]{Eggers:1995jq}
\bibinfo{author}{\bibfnamefont{H.~C.} \bibnamefont{Eggers}},
  \bibinfo{author}{\bibfnamefont{R.}~\bibnamefont{Tabti}},
  \bibinfo{author}{\bibfnamefont{C.}~\bibnamefont{Gale}}, \bibnamefont{and}
  \bibinfo{author}{\bibfnamefont{K.}~\bibnamefont{Haglin}},
  \bibinfo{journal}{Phys. Rev. D} \textbf{\bibinfo{volume}{53}},
  \bibinfo{pages}{4822} (\bibinfo{year}{1996}), \eprint{hep-ph/9510409}.

\bibitem[{\citenamefont{Sch\"afer et~al.}(2022)\citenamefont{Sch\"afer,
  Garcia-Montero, Paquet, Elfner, and Gale}}]{Schafer:2021slz}
\bibinfo{author}{\bibfnamefont{A.}~\bibnamefont{Sch\"afer}},
  \bibinfo{author}{\bibfnamefont{O.}~\bibnamefont{Garcia-Montero}},
  \bibinfo{author}{\bibfnamefont{J.-F.} \bibnamefont{Paquet}},
  \bibinfo{author}{\bibfnamefont{H.}~\bibnamefont{Elfner}}, \bibnamefont{and}
  \bibinfo{author}{\bibfnamefont{C.}~\bibnamefont{Gale}},
  \bibinfo{journal}{Phys. Rev. C} \textbf{\bibinfo{volume}{105}},
  \bibinfo{pages}{044910} (\bibinfo{year}{2022}), \eprint{2111.13603}.

\bibitem[{\citenamefont{Berges et~al.}(2017)\citenamefont{Berges, Reygers,
  Tanji, and Venugopalan}}]{Berges:2017eom}
\bibinfo{author}{\bibfnamefont{J.}~\bibnamefont{Berges}},
  \bibinfo{author}{\bibfnamefont{K.}~\bibnamefont{Reygers}},
  \bibinfo{author}{\bibfnamefont{N.}~\bibnamefont{Tanji}}, \bibnamefont{and}
  \bibinfo{author}{\bibfnamefont{R.}~\bibnamefont{Venugopalan}},
  \bibinfo{journal}{Phys. Rev. C} \textbf{\bibinfo{volume}{95}},
  \bibinfo{pages}{054904} (\bibinfo{year}{2017}), \eprint{1701.05064}.

\bibitem[{\citenamefont{Monnai}(2022)}]{Monnai:2022hfs}
\bibinfo{author}{\bibfnamefont{A.}~\bibnamefont{Monnai}},
  \bibinfo{journal}{Int. J. Mod. Phys. A} \textbf{\bibinfo{volume}{37}},
  \bibinfo{pages}{2230006} (\bibinfo{year}{2022}), \eprint{2203.13208}.

\bibitem[{\citenamefont{Monnai}(2020)}]{Monnai:2019vup}
\bibinfo{author}{\bibfnamefont{A.}~\bibnamefont{Monnai}}, \bibinfo{journal}{J.
  Phys. G} \textbf{\bibinfo{volume}{47}}, \bibinfo{pages}{075105}
  (\bibinfo{year}{2020}), \eprint{1907.09266}.

\bibitem[{\citenamefont{Churchill et~al.}(2021)\citenamefont{Churchill, Yan,
  Jeon, and Gale}}]{Churchill:2020uvk}
\bibinfo{author}{\bibfnamefont{J.}~\bibnamefont{Churchill}},
  \bibinfo{author}{\bibfnamefont{L.}~\bibnamefont{Yan}},
  \bibinfo{author}{\bibfnamefont{S.}~\bibnamefont{Jeon}}, \bibnamefont{and}
  \bibinfo{author}{\bibfnamefont{C.}~\bibnamefont{Gale}},
  \bibinfo{journal}{Phys. Rev. C} \textbf{\bibinfo{volume}{103}},
  \bibinfo{pages}{024904} (\bibinfo{year}{2021}), \eprint{2008.02902}.

\bibitem[{\citenamefont{van Hees et~al.}(2015)\citenamefont{van Hees, He, and
  Rapp}}]{vanHees:2014ida}
\bibinfo{author}{\bibfnamefont{H.}~\bibnamefont{van Hees}},
  \bibinfo{author}{\bibfnamefont{M.}~\bibnamefont{He}}, \bibnamefont{and}
  \bibinfo{author}{\bibfnamefont{R.}~\bibnamefont{Rapp}},
  \bibinfo{journal}{Nucl. Phys. A} \textbf{\bibinfo{volume}{933}},
  \bibinfo{pages}{256} (\bibinfo{year}{2015}), \eprint{1404.2846}.

\bibitem[{\citenamefont{van Hees et~al.}(2011)\citenamefont{van Hees, Gale, and
  Rapp}}]{vanHees:2011vb}
\bibinfo{author}{\bibfnamefont{H.}~\bibnamefont{van Hees}},
  \bibinfo{author}{\bibfnamefont{C.}~\bibnamefont{Gale}}, \bibnamefont{and}
  \bibinfo{author}{\bibfnamefont{R.}~\bibnamefont{Rapp}},
  \bibinfo{journal}{Phys. Rev. C} \textbf{\bibinfo{volume}{84}},
  \bibinfo{pages}{054906} (\bibinfo{year}{2011}), \eprint{1108.2131}.

\bibitem[{\citenamefont{Rapp}(2013)}]{Rapp:2013ema}
\bibinfo{author}{\bibfnamefont{R.}~\bibnamefont{Rapp}}, \bibinfo{journal}{PoS}
  \textbf{\bibinfo{volume}{CPOD2013}}, \bibinfo{pages}{008}
  (\bibinfo{year}{2013}), \eprint{1306.6394}.

\bibitem[{\citenamefont{Garcia-Montero
  et~al.}(2020)\citenamefont{Garcia-Montero, L\"oher, Mazeliauskas, Berges, and
  Reygers}}]{Garcia-Montero:2019kjk}
\bibinfo{author}{\bibfnamefont{O.}~\bibnamefont{Garcia-Montero}},
  \bibinfo{author}{\bibfnamefont{N.}~\bibnamefont{L\"oher}},
  \bibinfo{author}{\bibfnamefont{A.}~\bibnamefont{Mazeliauskas}},
  \bibinfo{author}{\bibfnamefont{J.}~\bibnamefont{Berges}}, \bibnamefont{and}
  \bibinfo{author}{\bibfnamefont{K.}~\bibnamefont{Reygers}},
  \bibinfo{journal}{Phys. Rev. C} \textbf{\bibinfo{volume}{102}},
  \bibinfo{pages}{024915} (\bibinfo{year}{2020}), \eprint{1909.12246}.

\bibitem[{\citenamefont{Fujii et~al.}(2022)\citenamefont{Fujii, Itakura,
  Miyachi, and Nonaka}}]{Fujii:2022hxa}
\bibinfo{author}{\bibfnamefont{H.}~\bibnamefont{Fujii}},
  \bibinfo{author}{\bibfnamefont{K.}~\bibnamefont{Itakura}},
  \bibinfo{author}{\bibfnamefont{K.}~\bibnamefont{Miyachi}}, \bibnamefont{and}
  \bibinfo{author}{\bibfnamefont{C.}~\bibnamefont{Nonaka}}
  (\bibinfo{year}{2022}), \eprint{2204.03116}.

\bibitem[{\citenamefont{Borsanyi et~al.}(2010)\citenamefont{Borsanyi, Endrodi,
  Fodor, Jakovac, Katz, Krieg, Ratti, and Szabo}}]{Borsanyi:2010cj}
\bibinfo{author}{\bibfnamefont{S.}~\bibnamefont{Borsanyi}},
  \bibinfo{author}{\bibfnamefont{G.}~\bibnamefont{Endrodi}},
  \bibinfo{author}{\bibfnamefont{Z.}~\bibnamefont{Fodor}},
  \bibinfo{author}{\bibfnamefont{A.}~\bibnamefont{Jakovac}},
  \bibinfo{author}{\bibfnamefont{S.~D.} \bibnamefont{Katz}},
  \bibinfo{author}{\bibfnamefont{S.}~\bibnamefont{Krieg}},
  \bibinfo{author}{\bibfnamefont{C.}~\bibnamefont{Ratti}}, \bibnamefont{and}
  \bibinfo{author}{\bibfnamefont{K.~K.} \bibnamefont{Szabo}},
  \bibinfo{journal}{JHEP} \textbf{\bibinfo{volume}{11}}, \bibinfo{pages}{077}
  (\bibinfo{year}{2010}), \eprint{1007.2580}.

\bibitem[{\citenamefont{Song and Coci}(2022)}]{Song:2021mvc}
\bibinfo{author}{\bibfnamefont{T.}~\bibnamefont{Song}} \bibnamefont{and}
  \bibinfo{author}{\bibfnamefont{G.}~\bibnamefont{Coci}},
  \bibinfo{journal}{Nucl. Phys. A} \textbf{\bibinfo{volume}{1028}},
  \bibinfo{pages}{122539} (\bibinfo{year}{2022}), \eprint{2104.10987}.

\bibitem[{\citenamefont{Song and Moreau}(2018)}]{Song:2018wvd}
\bibinfo{author}{\bibfnamefont{T.}~\bibnamefont{Song}} \bibnamefont{and}
  \bibinfo{author}{\bibfnamefont{P.}~\bibnamefont{Moreau}},
  \bibinfo{journal}{Phys. Rev. D} \textbf{\bibinfo{volume}{98}},
  \bibinfo{pages}{116007} (\bibinfo{year}{2018}), \eprint{1810.08013}.

\bibitem[{\citenamefont{Song et~al.}(2022)\citenamefont{Song, Grishmanovskii,
  and Soloveva}}]{Song:2022wil}
\bibinfo{author}{\bibfnamefont{T.}~\bibnamefont{Song}},
  \bibinfo{author}{\bibfnamefont{I.}~\bibnamefont{Grishmanovskii}},
  \bibnamefont{and} \bibinfo{author}{\bibfnamefont{O.}~\bibnamefont{Soloveva}}
  (\bibinfo{year}{2022}), \eprint{2210.04010}.

\bibitem[{\citenamefont{Low}(1958)}]{Low:1958sn}
\bibinfo{author}{\bibfnamefont{F.~E.} \bibnamefont{Low}},
  \bibinfo{journal}{Phys. Rev.} \textbf{\bibinfo{volume}{110}},
  \bibinfo{pages}{974} (\bibinfo{year}{1958}).

\bibitem[{\citenamefont{Koch et~al.}(1990)\citenamefont{Koch, Bl\"attel,
  Cassing, and Mosel}}]{Koch:1990jzd}
\bibinfo{author}{\bibfnamefont{V.}~\bibnamefont{Koch}},
  \bibinfo{author}{\bibfnamefont{B.}~\bibnamefont{Bl\"attel}},
  \bibinfo{author}{\bibfnamefont{W.}~\bibnamefont{Cassing}}, \bibnamefont{and}
  \bibinfo{author}{\bibfnamefont{U.}~\bibnamefont{Mosel}},
  \bibinfo{journal}{Phys. Lett. B} \textbf{\bibinfo{volume}{236}},
  \bibinfo{pages}{135} (\bibinfo{year}{1990}).

\bibitem[{\citenamefont{Peskin and Schroeder}(1995)}]{Peskin:1995ev}
\bibinfo{author}{\bibfnamefont{M.~E.} \bibnamefont{Peskin}} \bibnamefont{and}
  \bibinfo{author}{\bibfnamefont{D.~V.} \bibnamefont{Schroeder}},
  \emph{\bibinfo{title}{{An Introduction to quantum field theory}}}
  (\bibinfo{publisher}{Addison-Wesley}, \bibinfo{address}{Reading, USA},
  \bibinfo{year}{1995}), ISBN \bibinfo{isbn}{978-0-201-50397-5}.

\bibitem[{\citenamefont{Greco et~al.}(2003)\citenamefont{Greco, Ko, and
  Levai}}]{Greco:2003xt}
\bibinfo{author}{\bibfnamefont{V.}~\bibnamefont{Greco}},
  \bibinfo{author}{\bibfnamefont{C.~M.} \bibnamefont{Ko}}, \bibnamefont{and}
  \bibinfo{author}{\bibfnamefont{P.}~\bibnamefont{Levai}},
  \bibinfo{journal}{Phys. Rev. Lett.} \textbf{\bibinfo{volume}{90}},
  \bibinfo{pages}{202302} (\bibinfo{year}{2003}), \eprint{nucl-th/0301093}.

\bibitem[{\citenamefont{Andronic et~al.}(2021)\citenamefont{Andronic,
  Braun-Munzinger, Redlich, and Stachel}}]{Andronic:2021dkw}
\bibinfo{author}{\bibfnamefont{A.}~\bibnamefont{Andronic}},
  \bibinfo{author}{\bibfnamefont{P.}~\bibnamefont{Braun-Munzinger}},
  \bibinfo{author}{\bibfnamefont{K.}~\bibnamefont{Redlich}}, \bibnamefont{and}
  \bibinfo{author}{\bibfnamefont{J.}~\bibnamefont{Stachel}}, in
  \emph{\bibinfo{booktitle}{{Criticality in QCD and the Hadron Resonance Gas}}}
  (\bibinfo{year}{2021}), \eprint{2101.05747}.

\bibitem[{\citenamefont{Burnett and Kroll}(1968)}]{Burnett:1967km}
\bibinfo{author}{\bibfnamefont{T.~H.} \bibnamefont{Burnett}} \bibnamefont{and}
  \bibinfo{author}{\bibfnamefont{N.~M.} \bibnamefont{Kroll}},
  \bibinfo{journal}{Phys. Rev. Lett.} \textbf{\bibinfo{volume}{20}},
  \bibinfo{pages}{86} (\bibinfo{year}{1968}).

\bibitem[{\citenamefont{Aggarwal et~al.}(2000)}]{WA98:2000ulw}
\bibinfo{author}{\bibfnamefont{M.~M.} \bibnamefont{Aggarwal}}
  \bibnamefont{et~al.} (\bibinfo{collaboration}{WA98}) (\bibinfo{year}{2000}),
  \eprint{nucl-ex/0006007}.

\bibitem[{\citenamefont{Abreu et~al.}(2002)}]{NA50:2002edr}
\bibinfo{author}{\bibfnamefont{M.~C.} \bibnamefont{Abreu}} \bibnamefont{et~al.}
  (\bibinfo{collaboration}{NA50}), \bibinfo{journal}{Phys. Lett. B}
  \textbf{\bibinfo{volume}{530}}, \bibinfo{pages}{43} (\bibinfo{year}{2002}).

\bibitem[{\citenamefont{Song et~al.}(2011)\citenamefont{Song, Han, and
  Ko}}]{Song:2010fk}
\bibinfo{author}{\bibfnamefont{T.}~\bibnamefont{Song}},
  \bibinfo{author}{\bibfnamefont{K.~C.} \bibnamefont{Han}}, \bibnamefont{and}
  \bibinfo{author}{\bibfnamefont{C.~M.} \bibnamefont{Ko}},
  \bibinfo{journal}{Phys. Rev. C} \textbf{\bibinfo{volume}{83}},
  \bibinfo{pages}{024904} (\bibinfo{year}{2011}), \eprint{1012.0798}.

\bibitem[{\citenamefont{Adler et~al.}(2004)}]{PHENIX:2003iij}
\bibinfo{author}{\bibfnamefont{S.~S.} \bibnamefont{Adler}} \bibnamefont{et~al.}
  (\bibinfo{collaboration}{PHENIX}), \bibinfo{journal}{Phys. Rev. C}
  \textbf{\bibinfo{volume}{69}}, \bibinfo{pages}{034909}
  (\bibinfo{year}{2004}), \eprint{nucl-ex/0307022}.

\end{thebibliography}

\end{document}